\documentclass[
    aip,                 
    rsi,                 
    amsmath, amssymb,    
    preprint,            
    longbibliography     
]{revtex4-2}             
\usepackage{graphicx}     
\usepackage{dcolumn}      
\usepackage{bm}           
\usepackage[dvipsnames,svgnames,x11names]{xcolor}  
\usepackage[utf8]{inputenc}  
\usepackage[T1]{fontenc}  
\usepackage{mathptmx}     
\usepackage{setspace}     
\usepackage[colorlinks=true, citecolor=blue, urlcolor=blue, linkcolor=black]{hyperref} 
\usepackage{hypcap}       

\usepackage{placeins}     
\usepackage{etoolbox} 
\makeatletter
\patchcmd{\NAT@sort@cites}{\ifnum\NAT@type=\@ne\else}{}{}{}
\makeatother

\begin{document}
\doublespacing 

\title{Atomically Flat Dielectric Patterns for Band Gap Engineering and Lateral Junction Formation in MoSe$_2$ Monolayers} 

\author{Philipp Moser$^{+}$}
\affiliation{Walter Schottky Institute, Department of Physics, Technical University of Munich, 85748 Garching, Germany}
\affiliation{Physics Department, TUM School of Natural Sciences, Technical University of Munich, 85748 Garching, Germany}

\author{Lukas M. Wolz$^{+}$}
\affiliation{Walter Schottky Institute, Department of Physics, Technical University of Munich, 85748 Garching, Germany}
\affiliation{Physics Department, TUM School of Natural Sciences, Technical University of Munich, 85748 Garching, Germany}

\author{Alex Henning}
\affiliation{Walter Schottky Institute, Department of Physics, Technical University of Munich, 85748 Garching, Germany}
\affiliation{Physics Department, TUM School of Natural Sciences, Technical University of Munich, 85748 Garching, Germany}

\author{Andreas Thurn}
\affiliation{Walter Schottky Institute, Department of Physics, Technical University of Munich, 85748 Garching, Germany}
\affiliation{Physics Department, TUM School of Natural Sciences, Technical University of Munich, 85748 Garching, Germany}

\author{Matthias Kuhl}
\affiliation{Walter Schottky Institute, Department of Physics, Technical University of Munich, 85748 Garching, Germany}
\affiliation{Physics Department, TUM School of Natural Sciences, Technical University of Munich, 85748 Garching, Germany}

\author{Peirui Ji}
\affiliation{Walter Schottky Institute, Department of Physics, Technical University of Munich, 85748 Garching, Germany}
\affiliation{Physics Department, TUM School of Natural Sciences, Technical University of Munich, 85748 Garching, Germany}

\author{Pedro Soubelet}
\affiliation{Walter Schottky Institute, Department of Physics, Technical University of Munich, 85748 Garching, Germany}
\affiliation{Physics Department, TUM School of Natural Sciences, Technical University of Munich, 85748 Garching, Germany}

\author{Martin Schalk}
\affiliation{Walter Schottky Institute, Department of Physics, Technical University of Munich, 85748 Garching, Germany}
\affiliation{Physics Department, TUM School of Natural Sciences, Technical University of Munich, 85748 Garching, Germany}

\author{Johanna Eichhorn}
\affiliation{Physics Department, TUM School of Natural Sciences, Technical University of Munich, 85748 Garching, Germany}

\author{Ian D. Sharp}
\email[sharp@wsi.tum.de]{}
\affiliation{Walter Schottky Institute, Department of Physics, Technical University of Munich, 85748 Garching, Germany}
\affiliation{Physics Department, TUM School of Natural Sciences, Technical University of Munich, 85748 Garching, Germany}

\author{Andreas V. Stier}
\affiliation{Walter Schottky Institute, Department of Physics, Technical University of Munich, 85748 Garching, Germany}
\affiliation{Physics Department, TUM School of Natural Sciences, Technical University of Munich, 85748 Garching, Germany}

\author{Jonathan J. Finley}
\email[finley@wsi.tum.de]{}
\affiliation{Walter Schottky Institute, Department of Physics, Technical University of Munich, 85748 Garching, Germany}
\affiliation{Physics Department, TUM School of Natural Sciences, Technical University of Munich, 85748 Garching, Germany}

\begin{abstract}
Combining a precise sputter etching method with subsequent AlO$_x$ growth within an atomic layer deposition chamber enables fabrication of atomically flat lateral patterns of SiO$_2$ and AlO$_x$. The transfer of MoSe$_2$ monolayers onto these dielectrically modulated substrates results in formation of lateral heterojunctions, with the flat substrate topography leading to minimal strain across the junction. Kelvin probe force microscopy (KPFM) measurements show significant variations in the contact potential difference (CPD) across the interface, with AlO$_x$ regions inducing a 230~mV increase in CPD. Spatially resolved photoluminescence spectroscopy reveals shifts in spectral weight of neutral and charged exciton species across the different dielectric regions. On the AlO$_x$ side, the Fermi energy moves closer to the conduction band, leading to a higher trion-to-exciton ratio, indicating a bandgap shift consistent with CPD changes. In addition, transient reflection spectroscopy highlights the influence of the dielectric environment on carrier dynamics, with the SiO$_2$ side exhibiting rapid carrier decay typical of neutral exciton recombination. In contrast, the AlO$_x$ side shows slower, mixed decay behavior consistent with conversion of trions back into excitons. These results demonstrate how dielectric substrate engineering can tune the electronic and optical characteristics of proximal two-dimensional materials, allowing scalable fabrication of advanced junctions for novel (opto)electronics applications.
\end{abstract}


\maketitle 

\section{Introduction}\label{sec:intro}
Two-dimensional (2D) transition metal dichalcogenides (TMDs) are highly promising for the new generation of optoelectronic and quantum devices. They are direct bandgap semiconductors in the monolayer limit, have very high excitonic binding energies and novel valley and spin-dependent photonphysics.\cite{Mak10,Spl10,Wan12,But13,Xia14,Uge14} Moreover, when incorporated into devices, they can achieve high on/off current ratios in transistor structures and possess tunable optical and electronic properties that can be strongly modified by their proximity to other materials.  \cite{Rad11,Lop13,che14,Sti16,Yu16,Sti18,Liu21,Hu24} 
The development of scalable approaches for creating top-down integrated structures with nanoscale control over their optoelectronic properties would further widen their scope for applications.   \cite{Ros13,Fio14,Jar14,Chh16,Mov16,Wie17,Wol20} 
In this regard, the atomically thin nature of TMDs offers unique challenges and opportunities, with the supporting substrate and surrounding environment having a significant impact on their functional characteristics. For example, it is well known that the local dielectric environment surrounding such materials strongly influences their electronic and excitonic properties. \cite{Ryt67,Kel79,Kom12,Bra15,And15,Sti16b,Ryo16,Wie17,Nie19,Uta19,mhe24} 
In particular, the self-energy decreases with increasing dielectric constant, resulting in a redshift of the excitonic emission. However, the exciton binding energy is also reduced due to stronger screening, leading to an opposite blueshift. These effects are on the order of hundreds of meV and almost entirely cancel each other out. \cite{Uge14,Raj17,Qiu17,Cho18,Sch19,mhe24}
Nevertheless, the electronic band edges of the 2D materials experience significant shifts with changing dielectric surroundings, while electrostatic interactions with charged states in these layers can induce additional Fermi level shifts. These effects open up exciting possibilities for using intentionally inhomogeneous dielectric environments to imprint junctions and establish charge carrier variations within single flakes. \cite{Wie17,Uta19} \newline
In this context, atomic layer deposition (ALD) holds great potential for creating conformal coatings with atomically precise thickness and composition control. Indeed, ALD has been widely used to create advanced micro- and nano-scale optoelectronic devices, including those based on 2D materials. \cite{Gro04,Wel18,Gru22,Hen23,Hen23a} Among the broad range of dielectric coatings that can be produced by ALD, aluminum oxide (AlO$_x$) is a high-\( \kappa \) dielectric that is especially promising for low-temperature fabrication of protective layers on TMDs. Although such layers enable greatly enhanced durability, an understanding of their impact on the band edge energetics is still emerging. \cite{Nie19,Hen23}  Beyond providing a versatile approach for coating 2D materials, ALD can also be used to modify the dielectric properties of the underlying substrate. For example, our previous work on dielectric engineering and patterning using plasma-based ALD of AlO$_x$ on silicon dioxide (SiO$_2$) demonstrated the versatility of this technique for precise interface manipulation, yielding significant changes in local surface potential.\cite{Hen23a} \newline
In the present work, we introduce a new combination of sub-angstrom precise sputter etching with subsequent AlO$_x$ growth within the ALD chamber to achieve flat dielectric patterns that can be used to spatially modulate the properties of TMDs transferred onto them. In particular, we first sputter etched trenches in SiO$_2$ and then back-filled them with ALD AlO$_x$, both through a single predefined photoresist structure, to achieve laterally patterned dielectric surfaces with smooth interfaces between the AlO$_x$ and SiO$_2$ regions. The resulting substrates feature significant dielectric contrast between regions of alternating composition, with SiO$_2$ providing a relatively low dielectric constant, ranging from 3.8 to 3.9, and AlO$_x$ introducing a significantly larger value of up to $\sim$7.6. Here, it is important to note that the exact value for AlO$_x$ can vary depending on the growth process, film density, and thickness, thus offering further opportunities for tuning its dielectric properties.\cite{Gro02,Rob04,Gro04}
The utility of these structures for 2D/3D integration of TMDs is demonstrated by transferring a molybdenum diselenide (MoSe$_2$) monolayer across the interface, thus forming a lateral heterojunction within the TMD. While the dielectric contrast and resulting junction are comparable to that reported by Utama et al.\cite{Uta19}, our approach yields a seamless, step-free transition between different dielectric environments, thereby significantly reducing the impact of local strain compared with prior approaches. In addition, this technique ensures high reproducibility, can be applied at wafer scale, and maintains planar geometries that enhance compatibility with subsequent fabrication and deposition steps, providing a route to advanced dielectric-tuned optoelectronic devices. \newline
This study explores the optoelectronic properties of two distinct structures that span such ultra-flat and dielectrically patterned AlO$_x$/SiO$_2$ interfaces. The first is a transferred heterostructure composed of a MoSe$_2$ monolayer encapsulated on its top surface with hexagonal boron nitride (hBN) to enhance stability and optical quality. The second structure consists of an unencapsulated MoSe$_2$ monolayer transferred across the interface. This configuration facilitates the investigation of how the underlying substrate influences the work function of the TMD, junction formation, and molecular adsorption. We find that the MoSe$_2$ monolayers exhibit significant changes in optoelectronic properties across the dielectric interface. In particular, frequency-modulated Kelvin probe force microscopy (KPFM) measurements reveal a notable change in the contact potential difference (CPD) of over 230~meV, indicating the formation of a lateral heterojunction, spatially variable bandgap renormalization, and an interfacial electric field of 0.5~V/µm, which is expected to influence carrier dynamics and exciton dissociation. Indeed, low-temperature photoluminescence (PL) measurements on hBN-encapsulated flakes confirm these findings, with the ratio of the negatively charged trion (T) to neutral exciton (X$^0$) emission varying across the junction due to dielectric-tuned charge doping within the material. On the AlO$_x$ side, a higher trion spectral weight is observed, suggesting that the Fermi energy lies closer to the conduction band in this region. This shift in the Fermi energy with respect to the band edge positions significantly affects not only the optical but also the chemical properties of single MoSe$_2$ flakes. For the case of the unencapsulated monolayers, we find that adatoms readily attach to the TMD surface above the SiO$_2$, leading to strong defect PL emission (L-peak) at low temperatures. Conversely, such L-peak defect emission is absent from the AlO$_x$ region, suggesting reduced molecular adsorption. In addition, renormalization of the bandgap impacts carrier dynamics within the TMD. On the SiO$_2$ side, the neutral exciton exhibits the expected fast monoexponential decay.\cite{God16} In contrast, the AlO$_x$ side shows a bi-exponential decay, suggesting that trions decay into neutral excitons at high carrier densities. This process is more likely in regions where the Fermi energy is closer to the conduction band (i.e., above the AlO$_x$), explaining the differing lifetimes on the two sides of the dielectric interface. Overall, our work highlights the significant role of dielectric patterns in modifying the optoelectronic properties of TMDs, with the newly developed step-free structures facilitating junction formation with minimal influence from strain and providing a route to scalable 2D/3D integrated optoelectronic devices based on precise substrate engineering.

\section{Results and Discussion} \label{sec:results} 
\begin{figure*}
    \centering
     \includegraphics[trim={0mm 60mm 00mm 30mm},clip,width=.99\textwidth]{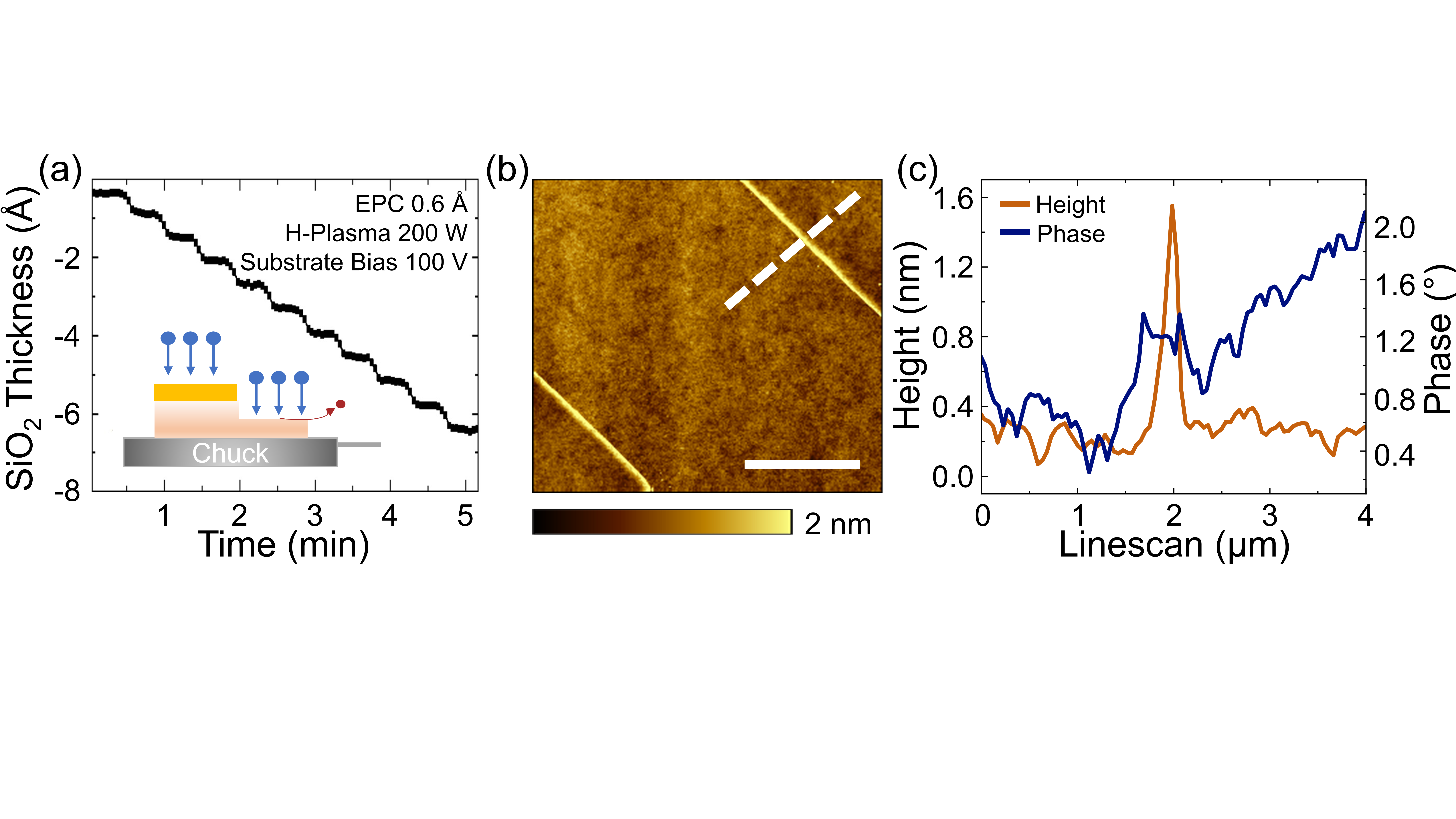} 
       \caption{\textbf{Etching process and surface analysis.} (a) Sputter etch rate per cycle (EPC), monitored on a reference SiO$_2$-coated Si substrate, demonstrating a precise EPC of 0.06~nm. Inset: Schematic of the etching method, highlighting the application of a rf bias to the chuck holder to accelerate H$^+$ ions from the H$_2$ plasma towards the substrate.      (b) Topography image of the surface after etching and AlO$_x$ deposition, showing a smooth transition between dielectrics with no step height and a small wall at the interface due to AlO$_x$ growth on the photoresist sidewalls. The scale bar represents a distance of 4~µm. (c) Line scan of the topography and phase across the dielectric pattern interface, analyzed along the white dashed line shown in (b).}
\label{fig:1}
\end{figure*}

To generate flat dielectric patterns, we developed a new process based on atomically precise sputtering of SiO$_2$ and subsequent ALD-based back-filling with AlO$_x$ through lithographically defined structures, as described in the Methods Section. In brief, lithographic resist patterns were generated on 270~nm thick SiO$_2$ films on n-type Si substrates, after which the samples were transferred to a plasma-enhanced ALD (PE-ALD) chamber equipped with a radio-frequency (rf) substrate bias chuck (Figure~S1(a)). To sputter the exposed regions of the sample with sub-angstrom level control, an rf bias was continuously applied to the substrate, and repetitive cycles of intermittent remote hydrogen plasma and gas purging were applied. During the sputtering steps, the substrate bias resulted in the acceleration of ionized species, primarily H$^+$ ions, towards the sample surface, as illustrated in the inset of \autoref{fig:1}(a). 

The process was conducted at a substrate temperature of 70~$^\circ$C and was carefully monitored using \textit{in situ} spectroscopic ellipsometry on a reference SiO$_2$-coated Si chip without photoresist, as shown in \autoref{fig:1}(a). The etch rate per cycle (EPC) can be adjusted by varying the bias voltage and the power of the H$_2$ plasma, as well as the plasma exposure time. Here, this controlled etching process was utilized to achieve an EPC of 0.06~nm/cycle for an rf substrate bias of 100~V, inductively coupled plasma power of 200~W, and plasma exposure time of 5~s. Etching cycles were performed to achieve a patterned trench depth of 3~nm, immediately after which 3~nm of AlO$_x$ was deposited using thermal ALD based on cyclic exposure of the sample to trimethylaluminum (TMA) and H$_2$O (see Methods Section for details). \cite{Gro04,Wel18} To precisely match the deposited thickness with the sputter etched thickness, the growth process was monitored via \textit{in situ} spectroscopic ellipsometry, as illustrated in Figure~S1(b), and was characterized by a growth per cycle (GPC) of 0.09~nm/cycle. 

After removing the sample from the ALD chamber, liftoff was performed, resulting in dielectric patterned substrates, as illustrated by the atomic force microscopy (AFM) image in \autoref{fig:1}(b).
\autoref{fig:1}(c) presents the line scan of the height across the interface between SiO$_2$ and AlO$_x$ regions, determined along the white dashed line indicated in \autoref{fig:1}(b). As intended, we observe almost no step height between the alternating dielectrics, though a narrow border wall of approximately 1.4~nm height is visible at the interface, likely due to AlO$_x$ growth on the photoresist sidewalls. Statistical analysis of the topography scan reveals that the average height difference between the AlO$_x$ side (top right) and the SiO$_2$ side (middle) is only 0.03~nm. Both sides are characterized by comparable root mean square roughnesses of 143~pm and 141~pm, respectively. Therefore, we find no measurable height difference between the two areas, as well as no significant change in the surface roughness. Along with the topographic profile, \autoref{fig:1}(c) illustrates the AFM phase change resulting from the different substrates. The evident phase shift between the AlO$_x$ and the SiO$_2$ regions is consistent with the successful fabrication of the alternating dielectric pattern.

The realization of a flat alternating oxide pattern provides a powerful basis for investigating the impact of different dielectric environments on proximal strain-free supported monolayers. Therefore, after fabricating the patterned substrate, we transferred two-dimensional MoSe$_2$ monolayers and heterostructures onto the dielectrically modulated substrate via mechanical exfoliation and viscoelastic stamping methods. \cite{Bah67,Wil69,Bro72,Nov04,Nov05} Here, MoSe$_2$ monolayers were chosen due to their well-understood electronic and optical properties, including energetically well-resolved exciton and trion emission. The exfoliated MoSe$_2$ monolayers were transferred onto the dielectric patterned substrates using viscoelastic stamping, with their top surfaces either encapsulated with hBN or bare, depending on the intended analysis. The individual dielectric regions of the substrate can be readily distinguished under an optical microscope, despite the lack of a physical step height between them. This allows for precise positioning of the flakes over the dielectric interface boundaries. For optical measurements, MoSe$_2$ flakes were encapsulated with hBN, while for electronic measurements based on KPFM, bare TMD flakes were transferred without hBN encapsulation. 

\begin{figure*}
    \centering
    \includegraphics[width=.99\textwidth]{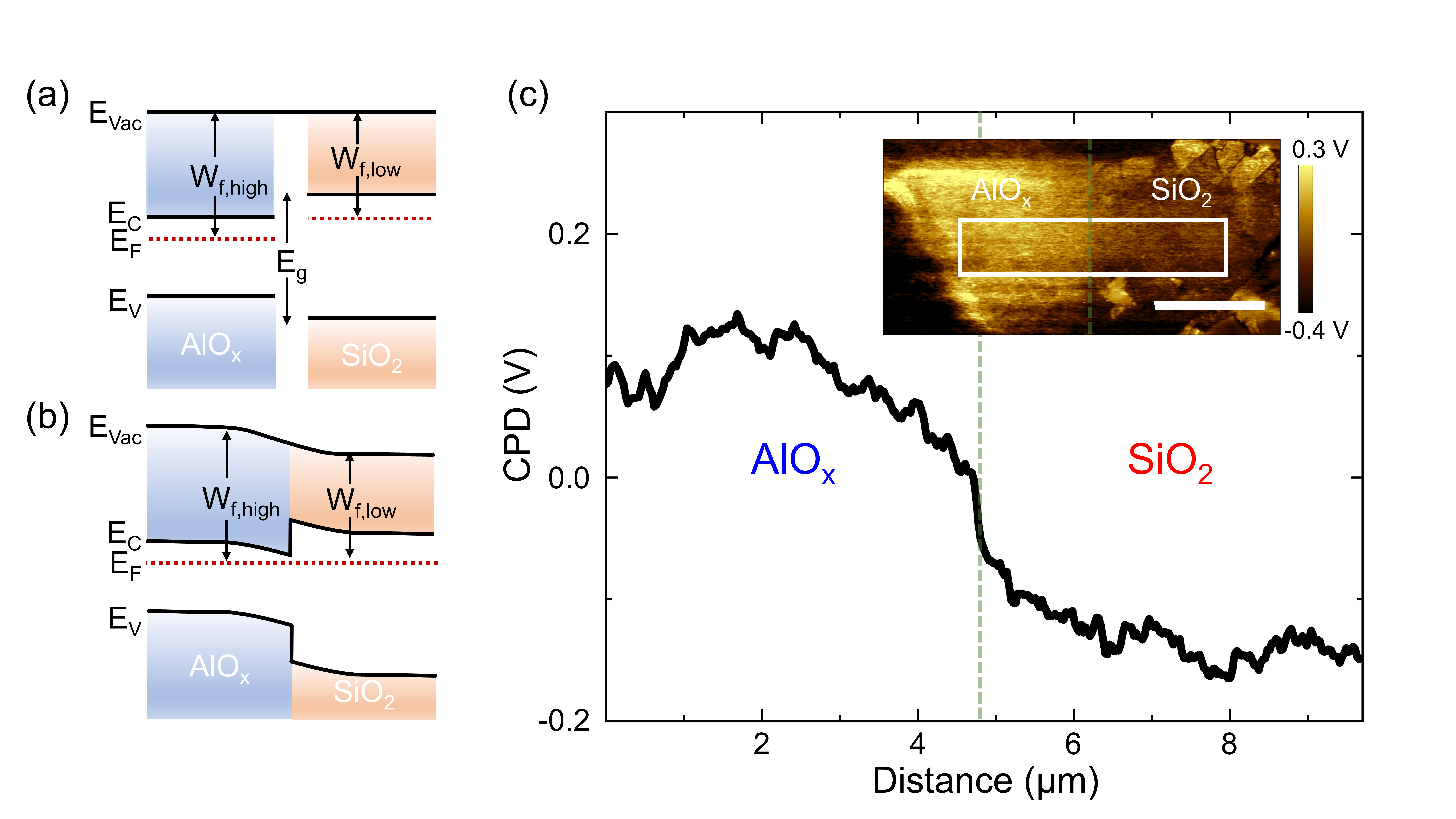}
    \caption{\textbf{Work function variations of MoSe$_2$ on a heterogeneous substrate.} (a) Schematic of bandgap renormalization in a MoSe$_2$ monolayer induced by differing underlying dielectric materials, highlighting predicted work function differences between AlO$_x$ and SiO$_2$ regions. (b) Predicted formation of a lateral heterojunction within the MoSe$_2$ monolayer due to spatially modified dielectric screening induced by the proximal interface between underlying AlO$_x$ and SiO$_2$ substrate regions. (c) Contact potential difference across an unencapsulated MoSe$_2$ monolayer spanning the interface between AlO$_x$ and SiO$_2$. The inset shows the spatially resolved CPD image of the MoSe$_2$ monolayer. The green dashed line marks the boundary between the substrates, and the white box marks the area used to extract the averaged CPD.}
\label{fig:2}
\end{figure*}

\autoref{fig:2}(a) illustrates schematically the expected impact of different local dielectric environments on the electronic structure of the MoSe$_2$ monolayer. The energy levels shown include the vacuum energy level ($E_{\text{vac}}$), conduction band minimum ($E_C$), valence band maximum ($E_V$), and the Fermi level ($E_F$). The energy bandgap ($E_g$) is indicated as the difference between the conduction band and valence band. The work functions, represented as $W_{f, \text{high}}$ and $W_{f, \text{low}}$, correspond to the MoSe$_2$ on AlO$_X$ (left) and SiO$_2$ (right), respectively. The region with the higher dielectric constant (AlO$_x$) is expected to undergo more significant bandgap renormalization than the side with the lower dielectric constant (SiO$_2$).\cite{Ryo16,Cho18} As such, we expect that the work function of the MoSe$_2$ monolayer on the AlO$_x$ side should shift to higher values due to greater renormalization compared to the work function on the SiO$_2$ side. These interactions of a single monolayer with contrasting dielectric regions of the substrate suggest that a lateral heterojunction will form above the AlO$_x$/SiO$_2$ interface, as shown in \autoref{fig:2}(b). \cite{Uta19} To test this hypothesis, we performed frequency-modulated KPFM measurements across the heterojunction, analyzing the local surface potential variations and directly measuring the band offset in the MoSe$_2$ heterojunction.
For these measurements, a higher CPD indicates a larger work function, as validated using a calibration sample composed of an Al-Si-Au interface. The KPFM scan shown in the inset of \autoref{fig:2}(c) reveals the CPD of an unencapsulated MoSe$_2$ monolayer positioned over the dielectric interface. The green dashed line marks the interface, and the white box highlights the region averaged in \autoref{fig:2}(c) to quantify the CPD. The left side of the flake overlays AlO$_x$, while its right side covers SiO$_2$. The CPD is 230~mV larger on the side with the higher dielectric constant, indicating a correspondingly higher work function, which aligns well with expectations based on the dielectric contrast.  
g
These observations strongly suggest the formation of a lateral heterojunction within the monolayer. Ideally, without trap states, the CPD would exhibit a sharp step function directly at the interface due to the abrupt change in dielectric properties. However, the observed electric field derived from the CPD across the interface reveals a more complex interaction. A distinct narrow electric field with a strength of at least 0.5~V/µm is created at the interface, as shown in Figure~S2, reflecting the immediate effect of the dielectric contrast. In the context of 2D materials like MoSe$_2$, a field strength of 0.5~V/µm can significantly impact the behavior of charge carriers and exciton dissociation, suggesting that these dielectric patterns can be used to manipulate charge separation, transport, and dynamics (see below). Away from the interface, the observed gradual change in the CPD over several micrometers is indicative of the formation of depletion regions in the monolayer on either side of the junction. The spatial extent of the depletion region on the AlO$_x$ side is larger than on the SiO$_2$ side. Assuming that the space charge width varies as the inverse square root of the charged impurity concentration, our results suggest a lower density of trapped states on the AlO$_x$ side. The combination of the larger work function on the AlO$_x$ side, the sharp potential step at the interface, and the gradual change of CPD away from the interface supports the conclusion that a lateral heterojunction is formed due to interaction with the dielectric pattern, as shown in \autoref{fig:2}(b). 

\begin{figure*}
    \centering
     \includegraphics[trim={0mm 20mm 0mm 10mm},width=.99\textwidth]{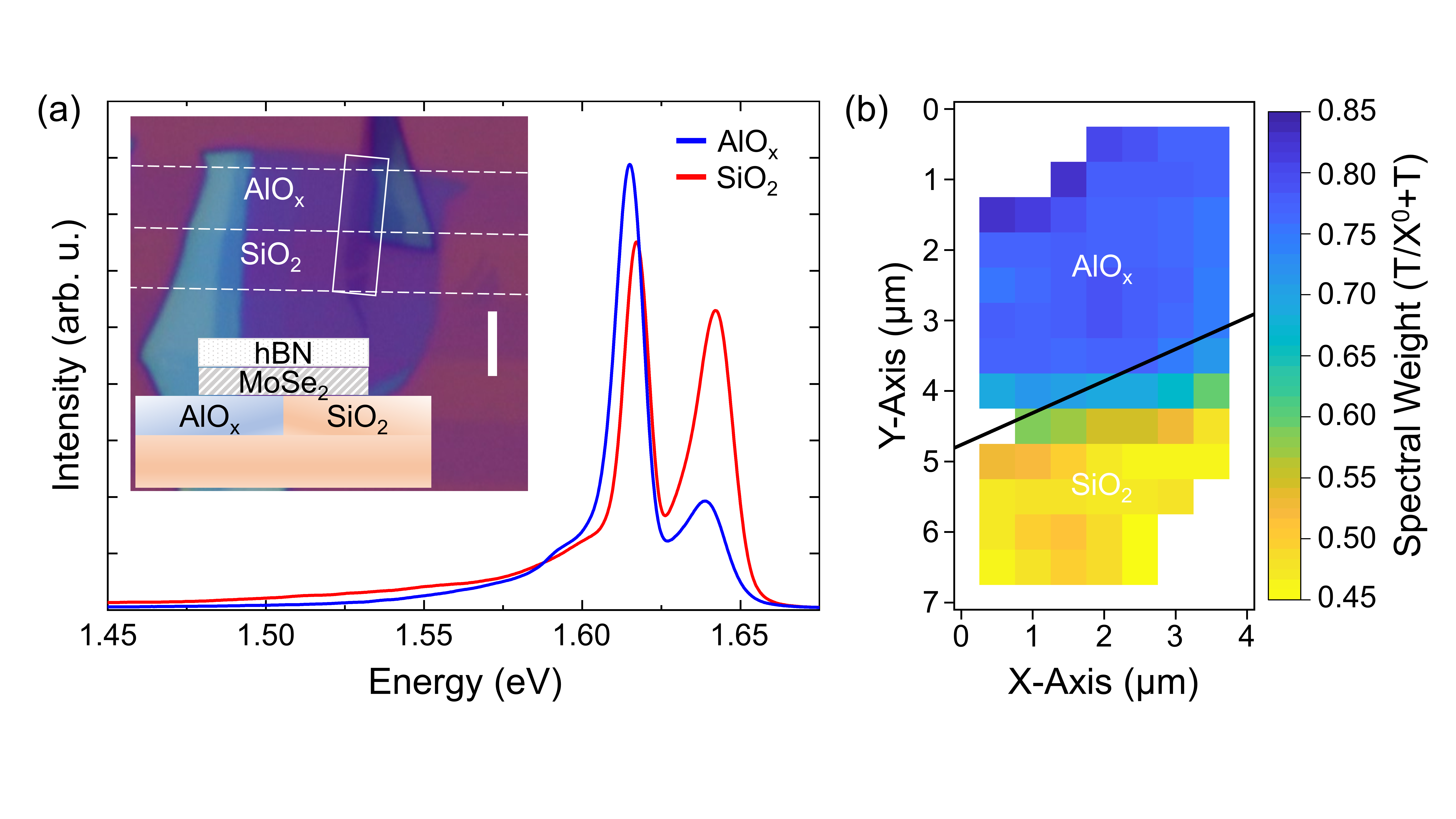}     
       \caption{\textbf{Impact of local dielectric environment on photoluminescence of MoSe$_2$.} (a) Low-temperature (10\,K) PL spectra of MoSe$_2$ placed onto the dielectric patterned substrate and covered by hBN, showing the neutral exciton (X$^0$) feature near 1.641~eV and trion (T) emission near 1.616~eV. The trion emission is notably more prominent over both AlO$_x$ (blue) and SiO$_2$ (red) regions. Inset: Image of the top-encapsulated MoSe$_2$ flake spanning the AlO$_x$/SiO$_2$ interface, with a scale bar indicating 5\,µm and dashed lines representing the boundaries between different dielectric regions, along with a schematic representation of a MoSe$_2$ monolayer placed across an interface of AlO$_x$ and SiO$_2$ regions, covered with hBN. The white box represents the measured area. (b) Spatial distribution of the spectral weight (T/(X$^0$+T)) across the flake, showing distinct values for the individual patterned regions, with the lateral interface indicated by the black line.}
\label{fig:3}
\end{figure*}

The low-temperature PL spectra of MoSe$_2$ with hBN encapsulation on a dielectric patterned substrate, obtained at 10 K, are shown in \autoref{fig:3}(a). A schematic and an optical image of the structure are shown in \autoref{fig:3}(a, inset). We note that encapsulated flakes are preferred for optical analysis since they provide narrower line widths and cleaner interfaces, thereby allowing changes due to the different substrates to be detected. For the encapsulated samples measured here, no defect-related PL emission is detected on either side of the dielectric interface, confirming the utility of the hBN. In addition, the excitonic peak at 1.641~eV is less intense than the trion emission at 1.616~eV in all locations, indicating n-type doping in the TMD. However, a more pronounced intensity difference and a slight red shift of approximately 3~meV are observed on the AlO$_x$ side. The greater n-type character in this region leads to a higher spectral weight from trion emission, as represented by the ratio of trion emission intensity to the total PL intensity of trions and neutral excitons (T/(X$^0$+T)). \autoref{fig:3}(b) further illustrates this variation of trion spectral weight as a function of position across the MoSe$_2$ flake spanning the lateral dielectric interface. On the AlO$_x$ side, the spectral weight (0.77) is larger and more uniform than on the SiO$_2$ side (0.55), which is fully consistent with the shift of the Fermi energy towards the conduction band due to increased dielectric screening by AlO$_x$. These observations are consistent with the band diagram depicted in \autoref{fig:2}(b). In particular, the combination of bandgap renormalization and decreased exciton binding energy due to enhanced screening shifts the Fermi energy closer to the conduction band edge. \cite{Gon13,Sim15,Uge14,Raj17,Qiu17,Cho18,Sch19,mhe24}  
Given the higher dielectric screening on the AlO$_x$ side, a reduced exciton binding energy is expected. While it is not straightforward to directly measure the exciton binding energy, the trion binding energy can be readily determined, allowing us to assess the role of the local dielectric environment on quasiparticle energetics. Assuming similar ratios for the binding energies of trions and excitons on both sides of the dielectric interface (i.e., due to each being similarly screened in the same local dielectric environment), the trion binding energy provides an effective indicator for the screening strength of the neutral exciton. Here, the trion binding energy is obtained from the energy difference between the neutral exciton and the trion photoluminescence lines, as obtained from the individual peak energies as a function of position shown in Figure~S3. 

\begin{figure*}
    \centering
     \includegraphics[trim={0mm 0mm 0mm 0mm},clip, width=0.99\textwidth]{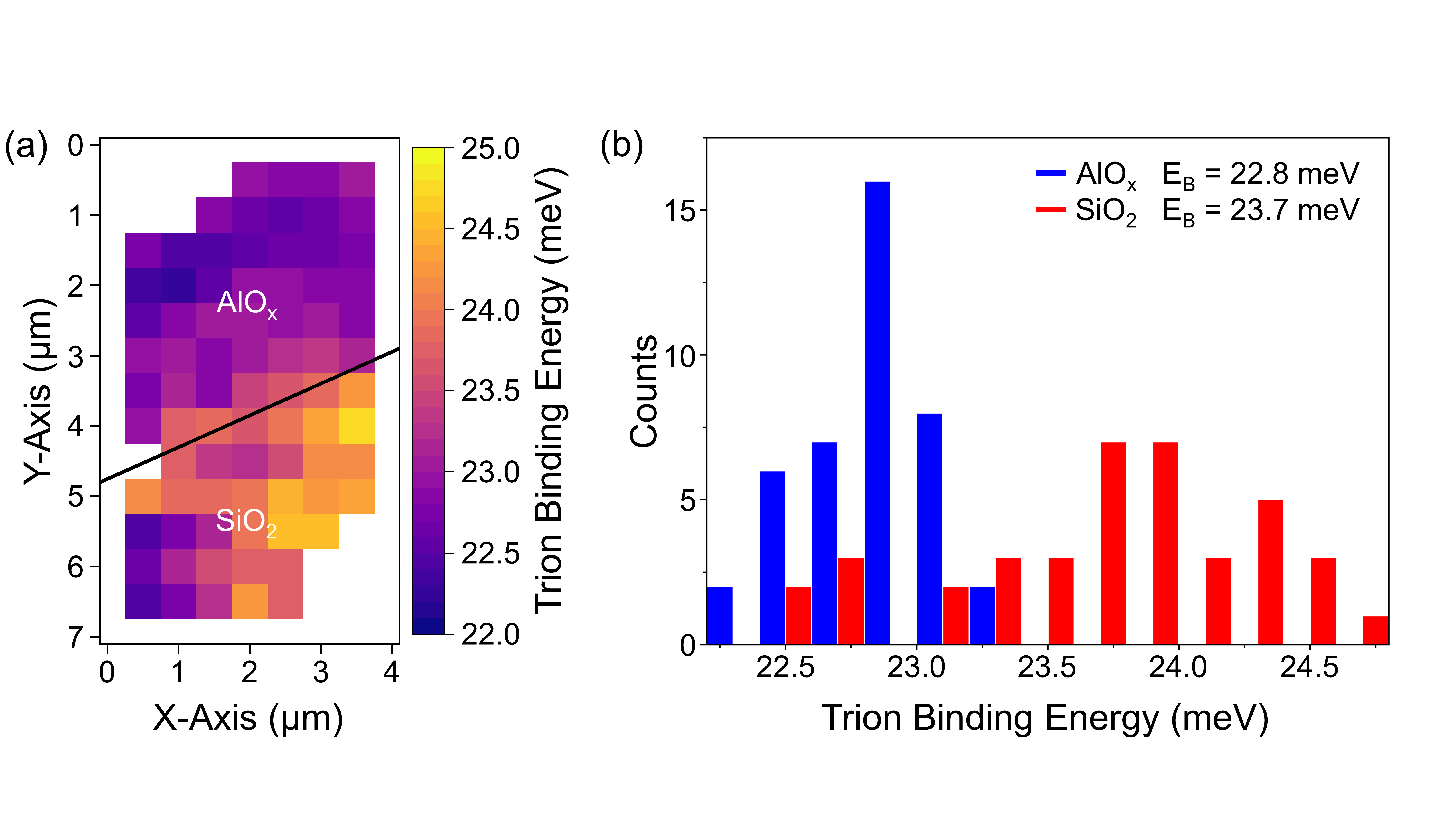} 
       \caption{\textbf{Trion binding energy distribution of a top-encapsulated MoSe$_2$ monolayer.} (a) Spatial map of the trion binding energy across a MoSe$_2$ monolayer situated over patterned AlO$_x$ and SiO$_2$ substrate regions, indicating the influence of underlying dielectric screening from the substrates. The black line indicates the lateral boundary between dielectric regions. (b) Histogram of trion binding energies obtained from the spatial map in (a), with a narrower distribution over AlO$_x$ (22.8~meV) compared to a broader distribution over SiO$_2$ (23.7~meV) suggesting different degrees of dielectric screening and quality of excitonic features between the two substrate regions.} \label{fig:4}
\end{figure*}

In \autoref{fig:4}(a), we present the trion binding energy as a function of position across the sample. It is characterized by a notable decrease on the AlO$_x$ side compared to the SiO$_2$ side. Indeed, the histogram in \autoref{fig:4}(b) reveals trion binding energies for the AlO$_x$ and SiO$_2$ side of 22.8~meV and 23.7~meV, respectively. This difference is consistent with increased screening due to the higher dielectric constant of AlO$_x$. The Keldysh-Rytova potential \( V(r) \), which is commonly used to describe the Coulomb interaction in 2D materials, can be used to quantify the change in exciton and trion binding energies over different regions of the dielectric pattern.\cite{Ryt67,Kel79,Ber13,Sti16b} Instead of experiencing a single dielectric constant, electrons and holes in atomically thin layers are influenced by the materials that are on their top and bottom surfaces, with relative dielectric constants given by \( \epsilon_t \) and \( \epsilon_b \), respectively. According to the Keldysh-Rytova formulation, \cite{Ryt67,Kel79} the potential \( V(r) \) is expressed as: 
\begin{equation}
V(r) = \frac{e^2}{8 \epsilon_0 r_0} \left[ H_0 \left( \frac{r \kappa}{r_0} \right) - Y_0 \left( \frac{r \kappa}{r_0} \right) \right]
\end{equation}
where \( H_0 \) and \( Y_0 \) are the Struve and Bessel functions, respectively. The screening length \( r_0 \) characterizes the dielectric properties of the TMD layer, while the average dielectric constant \( \kappa = (\epsilon_t + \epsilon_b)/2 \) accounts for the influence of the dielectric environment above and below the 2D material, with a larger \( \kappa \) resulting in more effective exciton screening. In accordance with previous studies,\cite{Sti18} we estimated the exciton binding energies in our MoSe$_2$ monolayer using this non-hydrogenic Keldysh-Rytova potential. For the dielectric constant on the AlO$_x$ side, we used the ratio of binding energies in the MoSe$_2$ monolayer transferred over an alternating AlO$_x$ and SiO$_2$ pattern, encapsulated with hBN. 
First, we calculated the exciton binding energy on the well-known SiO$_2$ side, using the dielectric constants of SiO$_2$ (\( \epsilon_{\text{SiO}_2} = 3.8 \)) and hBN (\( \epsilon_{\text{hBN}} = 4.5 \)), obtaining a value of 244~meV. By using the measured average trion binding energy of 23.7~meV on the SiO$_2$ side and scaling it by the ratio of the calculated exciton binding energies, we then estimated the exciton binding energy on the AlO$_x$ side to be 235~meV. Based on this value, we calculated the corresponding screening strength on the AlO$_x$ side, obtaining \( \epsilon_{\text{AlO}_X} = 4.12 \). As expected, this value is larger than on the SiO$_2$ side but is smaller than typically reported values for this material. However, it is important to recognize that the dielectric constant is underestimated due to differences in charge carrier densities in the different regions. In particular, the larger electron density on the AlO$_x$ side necessarily leads to an increase of the trion binding energy by a few meV, which partially offsets the decrease due to enhanced screening. Nevertheless, the observed change in trion binding energy unambiguously indicates increased screening, highlighting the significant impact of dielectric patterns on the optical properties of MoSe$_2$.

\begin{figure*}
    \centering
     \includegraphics[trim={0mm 65mm 0mm 20mm},clip, width=0.99\textwidth]{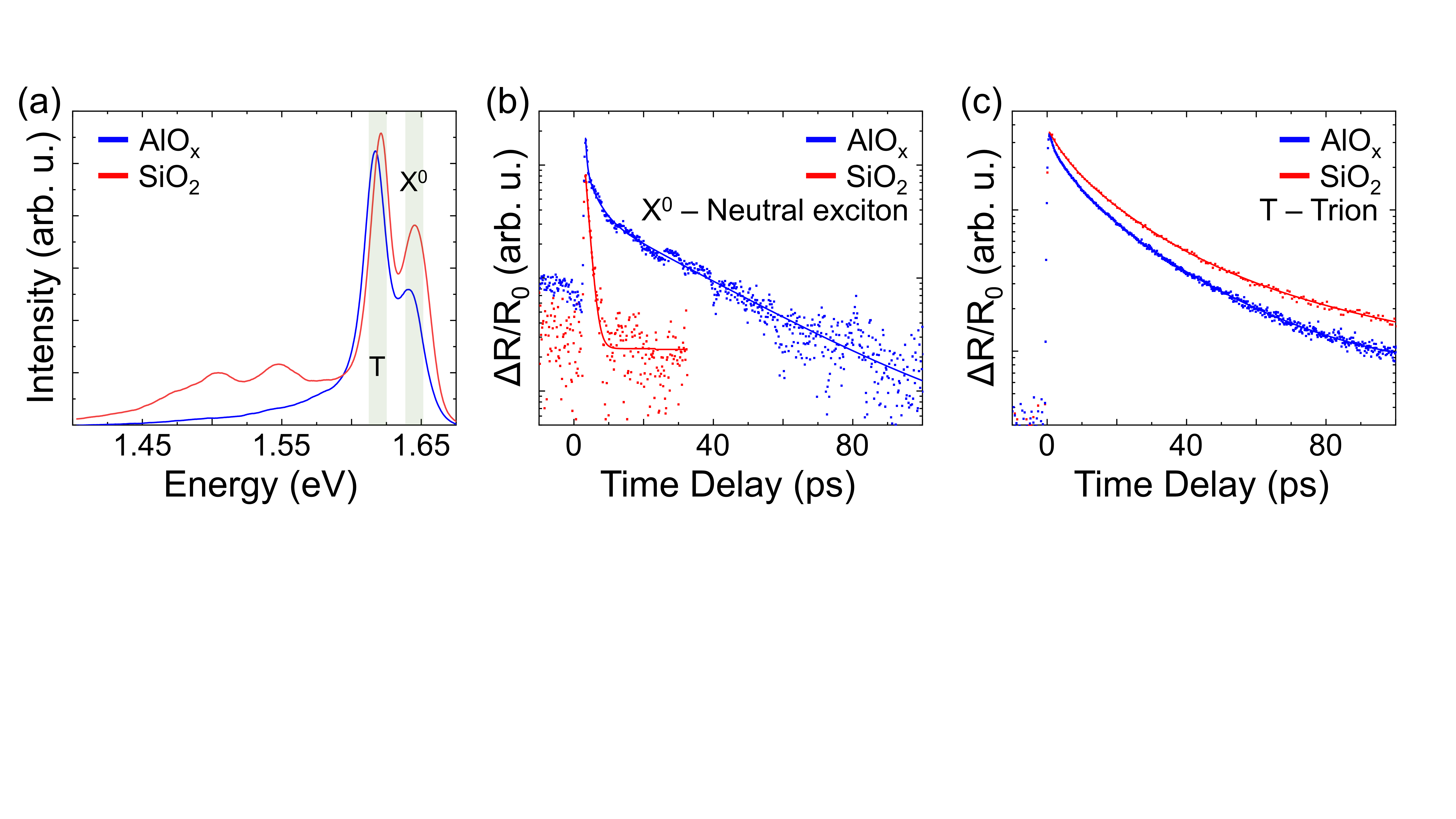} 
       \caption{\textbf{Impact of local dielectric environment on photoluminescence and carrier dynamics of MoSe$_2$.} (a) Low-temperature (10\,K) PL spectra of unencapsulated MoSe$_2$ placed across a lateral dielectric interface, measured in the AlO$_x$ (blue) and SiO$_2$ (red) regions. The regions probed during the transient reflection spectroscopy measurements are shaded in green. Transient reflection spectroscopy data of (b) the neutral exciton and (c) the negatively charged trion, measured at 10~K for MoSe$_2$ spanning the dielectric interface. } \label{fig:5}
\end{figure*}

The shift in Fermi energy across the interface not only changes the PL spectral weight and screening of exciton and trion binding energy but also influences other optical properties of the TMD, as well as its chemical interactions with the surroundings. These impacts are especially prominent for the case of unencapsulated MoSe$_2$, which can exhibit defect-associated L-peak emission due to molecular adsorption on the basal plane.\cite{Kle17,Wie17,Tan24} \autoref{fig:5}(a) shows low-temperature PL spectra recorded from an unencapsulated monolayer positioned over an alternating dielectric pattern. Importantly, the L-peak emission is significantly more pronounced on the SiO$_2$ side compared to the AlO$_x$ side, suggesting that the higher Fermi energy in the latter region inhibits molecular adsorption. Another possible explanation would be that the defects are passivated by being charged with an electron. Here, the lack of defect emission for the case of hBN top-encapsulated films (\autoref{fig:3}(a)) supports the conclusion that L-peak emission for unencapsulated material arises from adsorbates. This aligns with the KPFM measurements, also indicating a lower density of trap states over the AlO$_x$ region. Thus, such dielectric patterns provide a route to locally tune chemical interactions of considerable relevance, for example, in catalysis and sensing applications.

 The shift of the Fermi energy also has a significant effect on carrier dynamics within MoSe$_2$. The regions measured during the time-resolved measurements are shaded in green in \autoref{fig:5}(a). \autoref{fig:5}(b) shows the time-dependent reflection of the neutral exciton transition obtained from a micro-focused pump-probe experiment at 10 K. In this measurement, we monitored changes in the reflectance at different positions of the supported unencapsulated MoSe$_2$ flake following exposure to a 150~fs pump pulse at 3.29~eV double the energy of the neutral exciton. Reflectance measurements were collected after defined time delays using a probe beam with a bandwidth of approximately 13~meV to detect changes at the exciton resonance (1.645~eV).
Importantly, markedly different decay dynamics are observed in regions of the flake that are above SiO$_2$ compared to those above AlO$_x$. On the SiO$_2$ side, we observe a fast monoexponential decay that is typical of neutral exciton recombination at low temperatures. \cite{Vel15,Wan15,God16,Wan24,Ore24} In contrast, both a fast exponential and a significantly slower stretched exponential decay are evident on the AlO$_x$ side. While the fast decay component is similar on both sides of the dielectric interface and reflects the typical behavior of neutral exciton recombination in TMDs, the slow component on the AlO$_x$ side suggests the influence of charged trions on exciton dynamics. To provide additional insight into this behavior, we directly probed the trion dynamics at 1.615~eV, following excitation at 3.23~eV. As shown in \autoref{fig:5}(c), the transient reflection signal associated with trions exhibits stretched monoexponential decay dynamics for both AlO$_x$ and SiO$_2$ regions.

In contrast with the neutral exciton decay behavior, which is unaffected by changes of the carrier density, trion decay is sensitive to the carrier density, as the energy of the trion changes with varying density. Therefore, the stretched exponential trion decay behavior can be attributed to the dependence of its resonance frequency on carrier density. Furthermore, at a sufficiently high carrier density, a fraction of the trions can dissociate into neutral excitons via the release of an electron to achieve equilibrium. This reaction, which results in neutral exciton generation, prolongs the apparent exciton lifetime in regions that are more n-type, such as above AlO$_x$, yielding the additional slow relaxation component observed while in resonance with the neutral exciton. This demonstrates that the surrounding dielectric environment significantly influences carrier dynamics, unlocking new possibilities for tuning optoelectronic devices.

\FloatBarrier
\section{Conclusion and Outlook} \label{sec:conclusion}
This study demonstrates the fabrication of lateral heterojunctions within MoSe$_2$ monolayers using atomically flat dielectric patterns fabricated by precise sputter etching within an atomic layer deposition chamber followed by AlO$_x$ growth to backfill the sputtered trenches. Kelvin probe force microscopy on monolayers transferred across the dielectric pattern confirmed the formation of in-plane heterojunctions, revealing a significant change in the contact potential difference of over 230~mV due to the differing dielectric environments between regions. This shift highlights the pronounced impact of the local dielectric properties on the electronic characteristics of MoSe$_2$. Indeed, spatially resolved photoluminescence spectroscopy at 10~K revealed distinct variations in trion and exciton spectral weights, as well as shifts in their emission energies, across the different substrate regions. The higher trion spectral weight observed on the AlO$_x$ side indicates that the Fermi energy lies closer to the conduction band, supporting the findings from the KPFM measurements. Additionally, we used the Keldysh-Rytova potential to explain the changes of quasiparticle binding energies as a function of the local dielectric environment, with the trion binding energy reflecting the strength of dielectric screening. The reduced trion binding energy on the AlO$_x$ side compared to the SiO$_2$ side aligns with the expected increase in screening due to the higher dielectric constant of AlO$_x$. The effect of the dielectric environment on carrier dynamics was further examined by transient reflection spectroscopy, which showed a fast monoexponential neutral exciton decay on the SiO$_2$ side. In contrast, the AlO$_x$ side exhibited an additional stretched exponential decay component extending to much longer times. This change in the lifetime of the neutral exciton can be explained by the shift of the Fermi energy due to the influence of the substrate. The prolonged decay of the neutral exciton on the AlO$_x$ side is linked to the trion lifetime, suggesting that the neutral exciton's temporal dynamics are extended due to the conversion of trions into neutral excitons and free electrons. These findings underscore how local dielectric engineering can be used to spatially modify the electronic and optical properties of two-dimensional materials, paving the way for advanced heterostructures. In addition, the sharp potential step at the lateral dielectric interface provides prospects for forming nanoscale electric potential wells, thereby enabling reduced dimensionality of 2D systems to 1D or even 0D systems. This reduction in dimensionality could open up new possibilities for exploring novel physical phenomena and device applications.

\section{Experimental Details}
\noindent 
The fabrication process was designed to ensure minimal contamination of the monolayer and the substrate. The substrate comprised an n-doped silicon wafer with a 270~nm thermal oxide layer. To define lateral dielectric patterns, photolithography was first performed using a maskless aligner (Heidelberg Instruments, MLA100), which allows for flexible structure design. A double-layer resist of polydimethylglutarimide (PMGI) and ma-P 1205 was applied to ensure a clean surface for subsequent steps. The first layer of resist, PMGI, was spin-coated at a speed of 3000 rpm, resulting in a thickness of approximately 75~nm. This layer was designed to lift off cleanly later in the process. Before applying the second photoresist layer, the wafer was baked at 190 °C for 5 min. The second resist layer, ma-P 1205, was spin-coated at the same speed, resulting in an additional thickness of 500~nm. After baking at 100 °C for 3 min, selected areas of the resist were illuminated with an exposure of 170 mJ/cm$^2$ in the maskless aligner. The resist pattern was then developed by immersing the wafer in tetramethylammonium hydroxide for 15~s. Afterward, this structure was transferred to the ALD chamber (Fiji G2, Veeco, Figure~S1(a)). 

Atomically flat dielectric patterns were created through sub-angstrom precise sputtering within the ALD chamber and subsequent AlO$_x$ ALD growth. For sputtering, a 100~V rf bias was applied to the chuck holder, which accelerated ionized species, primarily H$^+$ ions, from a remotely generated hydrogen plasma towards the silicon substrate, as illustrated in the inset of \autoref{fig:1}(a). In the present work, a 200~W hydrogen plasma was used, with intermittent exposure times of 5~s, between which the chamber was purged to remove sputtered species and ensure precise thickness control.
The process was conducted at a substrate temperature of 70~$^\circ$C and was carefully monitored in real-time using \textit{in situ} spectroscopic ellipsometry on a reference silicon dioxide coated silicon chip without photoresist, as shown in \autoref{fig:1}(a). The etch rate per cycle (EPC) can be adjusted by varying the substrate bias voltage, the power of the H$_2$ plasma, and the plasma exposure time. In the present work, this controlled etching process yielded an EPC of 0.06~nm/cycle. It is noted that this process is only effective with a double-layer resist; single-layer photoresists tend to leave polymer residues on the chip carrier, inhibiting the sputtering process. The deposition of AlO$_x$ followed a well-established procedure involving cyclic exposure to trimethylaluminum and H$_2$O, between which the chamber was purged with Ar. \cite{Gro04,Wel18} As illustrated in Figure~S1(b), the growth process was also monitored by \textit{in situ} spectroscopic ellipsometry and exhibited a growth per cycle of 0.07~nm/cycle. Following deposition, liftoff of the dual-layer resist was achieved using N-methyl-2-pyrrolidone, complemented by q-tip cleaning to ensure thorough removal. Subsequently, the substrate was cleaned with acetone and isopropanol, after which it was blow-dried with nitrogen to remove any remaining solvents. \newline
Both hBN and MoSe$_2$ flakes were first exfoliated onto a 270~nm SiO$_2$ film on a Si wafer via the scotch tape method. The thicknesses of the exfoliated flakes were identified under a microscope by observing their contrast differences. Selected MoSe$_2$ monolayers and hBN/MoSe$_2$ heterostructures were transferred onto a patterned substrate using a home-built stacker via a polycarbonate-assisted transfer process. To enhance adhesion, the substrate temperature was raised to 80~$^\circ$C. The flakes were lifted in sequence, from the top layer to the bottom, and the polycarbonate was subsequently melted at 180~$^\circ$C to secure the flakes precisely in their intended locations on the substrate. After placement, the polycarbonate was dissolved with chloroform, ensuring the flakes remained cleanly adhered to the substrate.
 \newline
Topography and phase measurements were conducted in tapping mode on a Bruker Multimode V microscope. These AFM measurements were performed under ambient conditions using NSG30 AFM probes from TipsNano, featuring a nominal tip radius of 8~nm, a typical resonance frequency of 320 kHz, and a force constant of 40~N/m. The height images covered an area of 10 × 10~µm$^2$. Frequency-modulated Kelvin probe force microscopy was utilized to measure the contact potential difference between the probe tip and the sample surface. The measurements were conducted under ambient conditions using a Bruker Dimension Icon atomic force microscope equipped with SCM-PIT V2 Pt-Ir coated probes.  Ti/Au electrical contacts were connected at the edge of the monolayer to enable efficient charge transfer.\newline
Photoluminescence spectroscopy was performed on patterned MoSe$_2$ at cryogenic temperatures to characterize the emission properties. The experiments were conducted at 10~K using a Nd:YAG laser with an excitation energy of 2.33~eV. The laser was focused to a spot size of approximately 1~$\mu$m on the sample using a 100x objective. For the measurements, the excitation power was set to 3~$\mu$W and the integration time to 15~s to ensure sufficient signal acquisition.  \newline
Transient reflection spectroscopy was performed using an 81~MHz mode-locked Ti:Sapphire laser, which produced 150~fs pulses tunable across the 750-850~nm wavelength range. A segment of this laser output underwent second harmonic generation in a beta barium borate crystal to generate the pump beam. A delay stage was incorporated to adjust the path length of the probe pulse, allowing for precise control over the time delay between the pump and probe pulses. The pump beam was modulated using a chopper set to 9.731 kHz. For the MoSe$_2$ monolayers situated across a SiO$_2$/AlO$_x$ interface, the experiments were carried out to probe exciton (trion) transitions with a 1.645~eV (1.615~eV) probe beam and a 3.29~eV (3.23~eV) pump beam, calibrated to pump and probe powers of approximately 80~$\mu$W and 10~$\mu$W, respectively, under conditions of 10~K.
\newpage
\section{Data Availability Statement}
Data is available on request from the authors.
\section*{AUTHORS’ CONTRIBUTIONS}
P.M. and L.M.W. contributed equally to this work.
\section*{acknowledgement}
This work was supported by the Deutsche Forschungsgemeinschaft (DFG, German Research Foundation) through the TUM International Graduate School of Science and Engineering (IGSSE), project FEPChem2D (13.01), and by the DFG under Germany's Excellence Strategy – EXC 2089/1 – 390776260. J. J. Finley gratefully acknowledges the DFG for support via projects FI947-8 and SPP2244 (FI 947/7-1 and FI 947/7-2).

\bibliographystyle{achemso}
\bibliography{MAIN_Template}
\cleardoublepage


\begin{center}
    {\large \textbf{Supplementary information:}} \\[1ex] 
\end{center}    
    {\large \textbf{Atomically Flat Dielectric Patterns for Band Gap Engineering and Lateral Junction Formation in MoSe$_2$ Monolayers}} \\[1ex] 
\noindent Philipp Moser$^{+1,2}$, Lukas M. Wolz$^{+1,2}$, Alex Henning$^{1,2}$, Andreas Thurn$^{1,2}$, Matthias Kuhl$^{1,2}$, Peirui Ji$^{1,2}$, Pedro Soubelet$^{1,2}$, Martin Schalk$^{1,2}$, Johanna Eichhorn$^{2}$, Ian D. Sharp$^{1,2,~a)}$, Andreas V. Stier$^{1,2}$, Jonathan J. Finley$^{1,2,~b)}$ \\[3ex]
    {\itshape
    \noindent $^1$Walter Schottky Institute, Department of Physics, Technical University of Munich, 85748 \newline Garching, Germany \\
    $^2$Physics Department, TUM School of Natural Sciences, Technical University of Munich, 85748 Garching, Germany} \\[1ex]

\setcounter{figure}{0} 
\renewcommand{\thefigure}{S\arabic{figure}} 
\FloatBarrier
\vspace{4.5cm}
\begin{figure}[ht]
    \centering
    \includegraphics[trim=0mm 50mm 0mm 70mm,width=\textwidth]{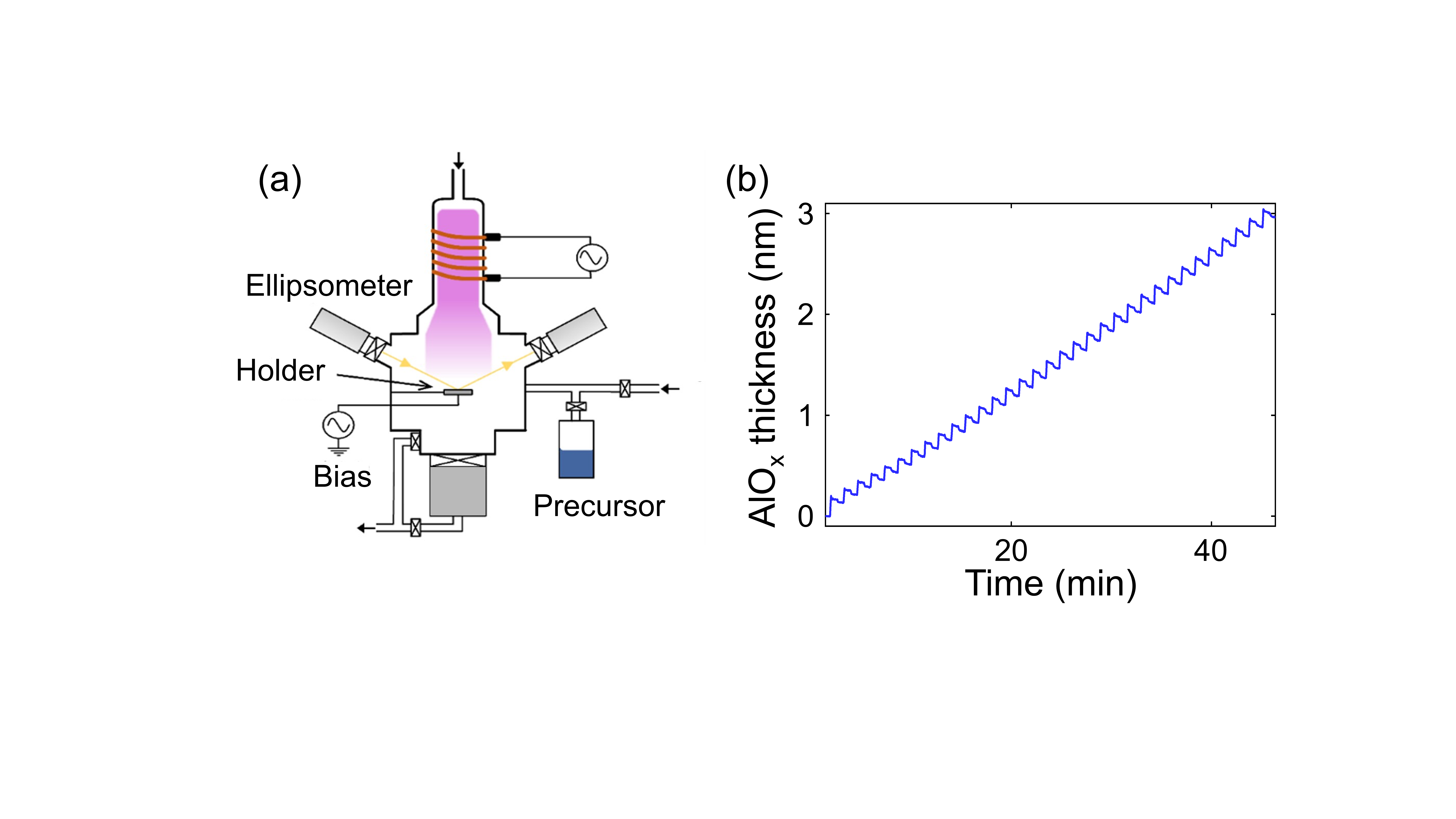}
    \caption{(a) Schematic of the ALD chamber, including the \textit{in situ} ellipsometer, remote plasma source, and rf substrate bias chuck. (b) Thickness of ALD AlO$_x$ layer as a function of time, monitored by \textit{in situ} spectroscopic ellipsometry.}
    \label{fig:S1}
\end{figure}

\begin{figure}
    \centering
    \includegraphics[trim=00mm 10mm 00mm 10mm,width=0.99\textwidth]{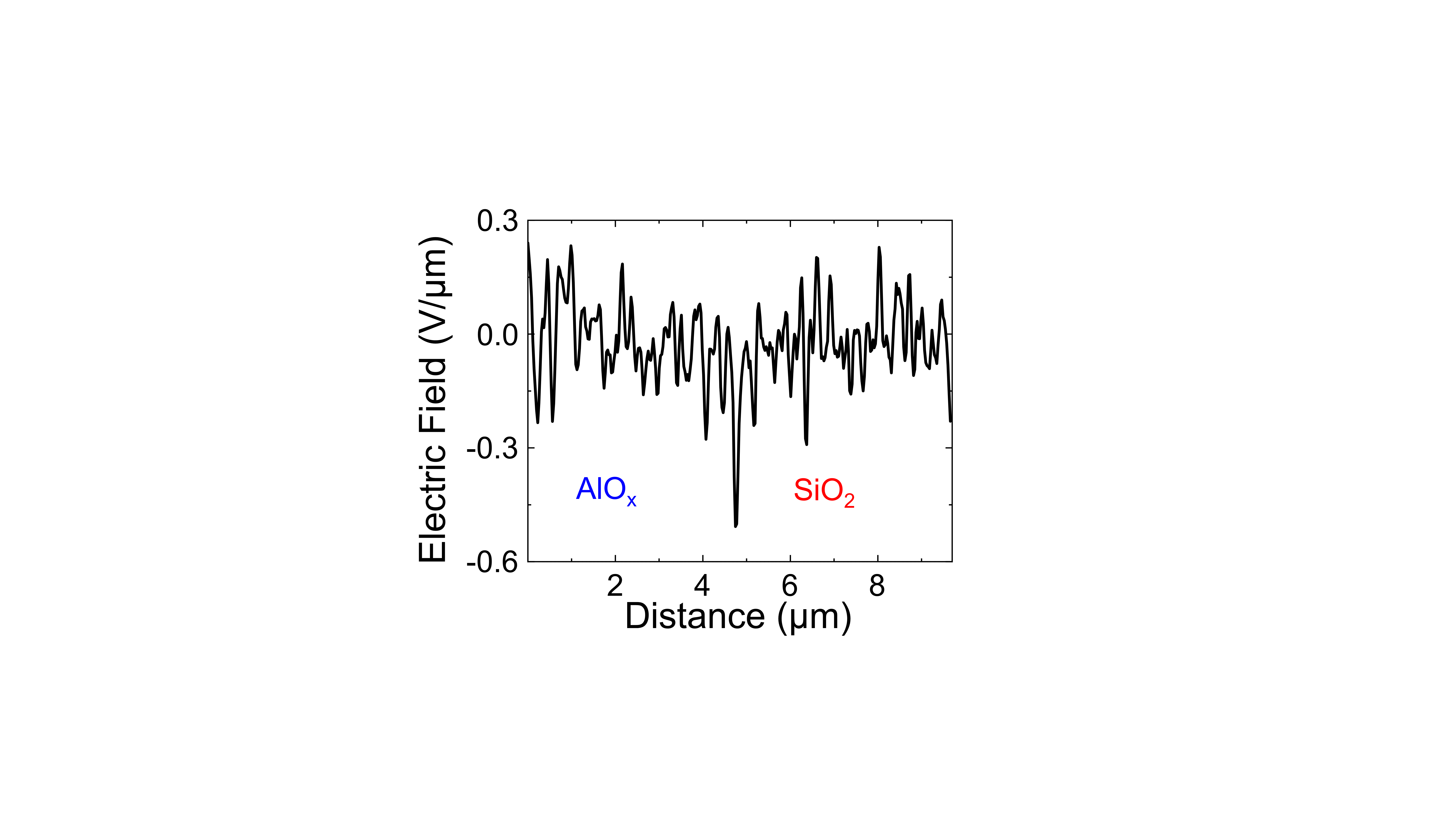}
    \caption{Electric field across the highlighted region in Figure 2(c), obtained from the derivative of the CPD signal with respect to position.}
    \label{fig:S2}
\end{figure}

\begin{figure} 
     \includegraphics[trim=0mm 0mm 0mm 0mm,width=0.99\textwidth]{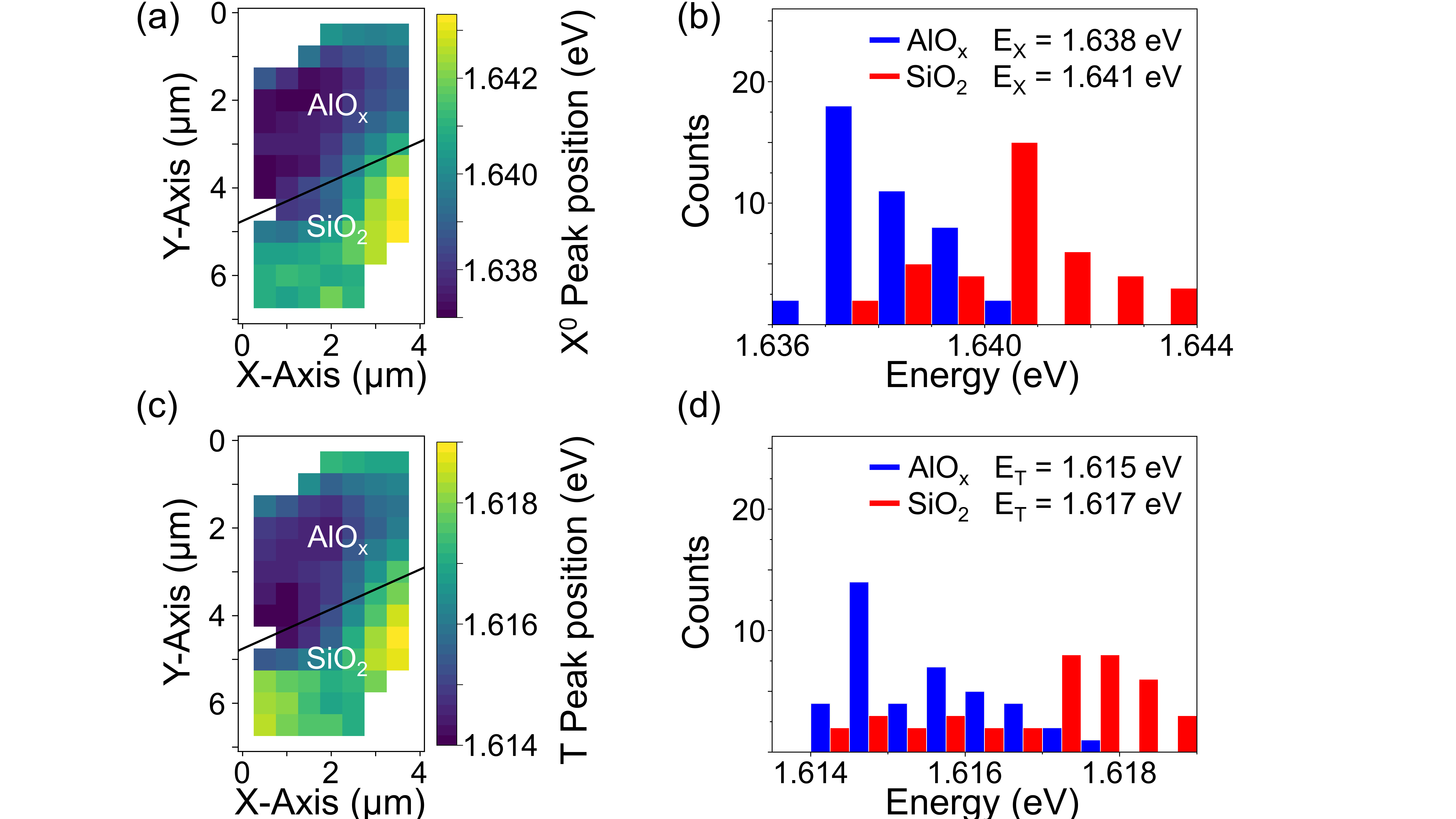} 
       \caption{(a) Spatial mapping of the neutral exciton peak position from a MoSe$_2$ monolayer placed across the lateral dielectric interface.
(b) Histogram analysis of the neutral exciton peak position from (a), obtained from the two dielectric regions.
(c) Spatial mapping of the negatively charged trion position across a MoSe$_2$ monolayer situated over alternating AlO$_X$ and SiO$_2$ regions.
(d) Histogram analysis of the trion peak position from (c), obtained from the two dielectric regions.}
 \label{fig:S3}
\end{figure}

\cleardoublepage

\end{document}             